# Photonic defect modes in cholesteric liquid crystal with a resonant nanocomposite layer and a twist defect

Stepan Ya. Vetrov[1,2], Maxim V. Pyatnov[1,2,3,*], and Ivan V. Timofeev[1,2]

[1]*L.V. Kirensky Institute of Physics, Siberian Branch of the Russian Academy of Sciences, Krasnoyarsk 660036, Russia*
[2]*Siberian Federal University, Krasnoyarsk 660041, Russia*
[3]*Krasnoyarsk State Medical University named after Professor V.F. Voyno-Yasenetsky, Krasnoyarsk 660022, Russia*
*Corresponding author: MaksPyatnov@yandex.ru*

We have studied spectral properties of a cholesteric liquid crystal with a combined defect consisting of a nanocomposite layer and a twist. The nanocomposite layer is made of metallic nanoballs dispersed in a transparent matrix and featuring effective resonant permittivity. A solution has been found for the transmission spectrum of circularly polarized waves in the structure. We have analyzed spectral splitting of the defect mode in the bandgap of the cholesteric when its frequency coincides with the nanocomposite resonant frequency. Defect modes have characteristics strongly dependent on the magnitude and direction of the phase jump at the interface between nanocomposite and cholesteric layers. It has been found that the bandgap width and the position and localization degree of defect modes can be effectively controlled by external fields applied to cholesteric.

## I. INTRODUCTION

Photonic crystals (PC) attract much interest due to their special optical properties and versatile application potential [1-4]. Such materials are unique in that their dielectric permittivity can change periodically in one, two or three dimensions at a spatial scale comparable with the light wavelength. There is a fairly close formal analogy between the theory of propagation of electromagnetic waves in periodic media and the quantum theory of electrons in crystals. The zone structure of electron energy spectrum associated with Bragg reflection of electrons is similar to that of a PC. In photonic bandgaps (PBG) of PCs with a lattice defect, i.e. with a disturbed periodicity, there emerge transmission bands with the location and transmission coefficient controllable by geometrical and structural parameters. Note that light is localized in the defect area, which enhances the light wave intensity inside the defect layer.

Cholesteric liquid crystals (CLC) belong to a special class of one-dimensional PCs possessing unique properties, namely broad transparency band, strong nonlinearity and high susceptibility to external low-frequency electric or magnetic fields as well as to applied light or heat [5,6]. The CLC helix pitch and location of the bandgap can be essentially changed by varying temperature and pressure and applying electromagnetic fields or mechanical stress. The qualitative difference of CLC from any other type of PC lies in their polarization selectivity by diffractive reflection. In CLCs there are PBGs for light propagating along the CLC helix with a circular polarization coinciding with the twist of the CLC helix. When reflected from CLC, a light wave of such polarization does not change the sign of its polarization.

Bragg reflection occurs in the wavelength range between $\lambda_1 = pn_o$ and $\lambda_2 = pn_e$, where $p$ is the helix pitch, $n_o$ and $n_e$ are the ordinary and extraordinary refractive indexes of CLC. Light of opposite circular polarization does not undergo diffractive reflection and passes through a CLC medium almost unaffected.

Introducing various types of defects into the structure of an ideal cholesteric makes it possible to obtain narrow transmission bands in PC bandgaps that will correspond to localized defect modes. These modes, similarly to defect modes in scalar periodic layered media, can be used to produce narrowband and tunable filters [7], optical diodes [8], liquid crystal rotator and a tunable polarizer [9], polarization azimuth rotation and azimuth stabilizers [10] and other devices. Low threshold laser generation in CLC is under extensive research [11]. There are two possible options of such generation: band edge lasing [12] and defect mode lasing [13].

There exist a few ways of inducing photonic defect modes in CLC. This can be achieved by using a thin layer of isotropic [14] or anisotropic substance [13,15] confined between two layers of CLC, twist defect (jump of the rotation angle of the cholesteric helix) or defect associated with local changes in the helix pitch [19-21]. The authors of [22], using an analytical approach to the theory of optical defect modes in CLC with an isotropic defect layer, developed a model that isolates polarization mixing and yields light equation only for diffractive polarization. Analysis of defect modes in CLC induced by twist defect, i.e. by a sharp turn of the director around the CLC axis at the interface of two CLC layers, can be found in [23,24]. The direction of predominant orientation of molecules in CLC is referred to as a CLC director. Propagation of light in CLC containing a dielectric layer combined with twist-defect at the interface between the dielectric and CLC layer was studied in [7,25]. PCs doped with substances having strong optical resonance are under extensive research. An example of a giant optical resonance has been predicted for a nanocomposite (NC) consisting of metallic nanoballs dispersed in a transparent matrix [26] while the optical characteristics of primary materials lack resonant properties. The resonance is characterized by effective permittivity in approximation of the Maxwell Garnett formula, which is proved by both the numerical simulation [27] and experiment [28]. The effective characteristics of a NC containing metallic nanoparticles dispersed in a dielectric matrix are due to the plasmon resonance of nanoparticles. These characteristics can have unique magnitudes not inherent in natural materials. For instance, the effective refractive index can be in excess of 10. Dispersion of resonant medium combined with the PC dispersion provides a new tool to control spectral and optical properties of PCs [29-31].

Our study is focused on spectral properties of CLC with a combined defect. A NC layer with resonance dispersion and a phase jump at the interface between the NC and CLC layers serve as such a defect. Splitting of the defect mode and localization of the electromagnetic field are analyzed depending on the volume fraction of nanoparticles in the defect layer. Phase jump of the CLC helix in the transmission spectrum is discussed. We also study modification of the transmission spectrum of cholesteric with a combined defect under external fields affecting the CLC helix pitch.

## II.  MODEL AND TRANSMITTANCE

The PC structure under consideration consists of two identical layers of CLC with a right-handed twist separated by a NC defect layer combined with a twist defect (Fig.1). The length of the entire structure is $L = 2.946$ µm, the CLC helix pitch is $p = 275$ nm, and the defect layer thickness is $d = 196$ nm. The medium outside CLC is isotropic and has the refractive index $n = (n_o+n_e)/2$, where $n_o = 1.4$ and $n_e = 1.6$ are the ordinary and extraordinary refractive indexes of cholesteric, respectively. For the chosen external medium, Fresnel reflection from the CLC surface and interference processes from the boundary surfaces will be weak. The twist defect magnitude is governed by the angle α that is positive when the vector tip of the CLC director rotates clockwise at the interface between NC and CLC layers (conventionally observed along the direction of light propagation).

The effective permittivity of a NC layer $\varepsilon_{mix}$ is found from the Maxwell Garnett formula widely used when dealing with matrix media where a small volume fraction of isolated inclusions is dispersed in the material matrix [26,32]:

$$\varepsilon_{mix} = \varepsilon_d \left[ \frac{f}{(1-f)/3 + \varepsilon_d/(\varepsilon_m - \varepsilon_d)} + 1 \right]. \tag{1}$$

Here $f$ is the filling factor, i.e. the fraction of nanoparticles in the matrix, $\varepsilon_d$ and $\varepsilon_m(\omega)$ are the permittivities of the matrix and the metal of nanoparticles, respectively, $\omega$ is the radiation frequency. The size of nanoparticles is much smaller than the wavelength and the depth of field penetration into the material. Find permittivity of the metal of nanoparticles using the Drude approximation:

$$\varepsilon_m(\omega) = \varepsilon_0 - \frac{\omega_p^2}{\omega(\omega + i\gamma)}, \quad (2)$$

where $\varepsilon_0$ is the background dielectric constant taking into account contributions from interband transitions, $\omega_p$ is the plasma frequency, $\gamma$ is the plasma relaxation rate . For silver nanoballs dispersed in a transparent optical glass, $\varepsilon_0 = 5$, $\hbar\omega_p = 9$ eV, $\hbar\gamma = 0.02$ eV, $\varepsilon_d = 2.56$.

Ignoring the small factor $\gamma^2$, we find position of the resonant frequency that depends on the characteristics of primary materials and the dispersed phase concentration $f$.

$$\omega_0 = \omega_p \sqrt{\frac{1-f}{3\varepsilon_d + (1-f)(\varepsilon_0 - \varepsilon_d)}}. \quad (3)$$

Figure 2 shows dispersion dependence of the nanocomposite permittivity for two different filling factors: $f = 0.02, 0.1$. As can be seen in the figure, the $\omega_0$ frequency corresponding to resonance in the defect layer shifts towards the long-wavelength range of the spectrum with the growing concentration of nanoballs. Note that the resonant curve half width $\varepsilon''_{mix}$ is very little affected while the $\varepsilon'_{mix}$ curve is essentially modified and the range of frequencies increases for which the NC is similar to metal when $\varepsilon'_{mix} < 0$. We employed the Berreman transfer matrix method for numerical analysis of the spectral properties and field distribution in the sample [33,34]. This provided us with a tool to numerically assess propagation of light in cholesteric with a structural defect. An equation to describe light propagation at frequency $\omega$ along the $z$ axis is given by

$$\frac{d\Psi}{dz} = \frac{i\omega}{c}\Delta(z)\Psi(z), \quad (4)$$

where $\Psi(z) = (E_x, H_y, E_y, -H_x)^T$, and $\Delta(z)$ is the Berreman matrix dependent on the permeability tensor and incident wave vector.

### III. RESULTS AND DISCUSSION

#### A. Optics of CLC with a combined defect

Figure 3a shows a seed ($f = 0$) transmission spectrum for normal incidence of light on CLC with a structural defect in the form of a dielectric platelet without twist defect ($\alpha = 0$). One can see that similarly to [14], there are peaks in the CLC PBG associated with CLC defect modes induced for both circular polarizations of normally incident light. Moreover, defect modes are associated with the same wavelength and the same transmission at the defect mode wavelength.

In [31] it was found that, similarly to frequency splitting of two coupled oscillators, the defect mode frequency splits if the filling factor is nonzero and the NC resonant frequency $\omega_0$ is close to the defect mode frequency. The effect of splitting in the transmission spectrum is illustrated in Fig.3b. As can be seen in the figure, the resultant defect modes have the same wavelength for the right- and left-handed polarizations but differ in transmittance in the peak centre. Calculations show that the splitting enhances with the growing volume fraction of nanoballs in the composite as it happens in the case of a scalar one-dimensional PC with a resonant defect nanocomposite layer [29]. The reflection index and the absorption coefficient were found to be strongly dependent on the direction of circular polarization of incident light. The emergent bandgap in the spectrum for both polarizations is primarily due to substantial reflection and absorption of waves with right-handed and left-handed polarizations, respectively.

The result of the twist-defect is the repositioning of defect modes in the bandgap of CLC, which affects transmission at the defect mode frequency. Figure 4a shows that clockwise rotation of the second CLC layer ($\alpha > 0$) makes both peaks corresponding to defect modes for right-handed and left-handed polarizations shift to the shorter wavelength range. Transmittance of the long-wavelength peak drops down with the growing $\alpha$, whereas that of the short-wavelength peak enhances. A reverse situation is observed when the CLC layer is rotated anti-clockwise ($\alpha < 0$) (Fig. 4b). Redistribution of transmission intensity at defect modes depending on $\alpha$ is obviously associated with their coupling due to the twist defect. No defect modes of either polarization emerge in the bandgap zone of CLC for the chosen parameters of the system and $\alpha = 90^0$. Our calculations show that peaks for light of non-diffractive polarization are not observed just for any $\alpha$. The bandgap location is almost independent of the twist defect magnitude.

Figure 5 is an example of spatial distribution of electric field in $\lambda = 388$ nm defect modes for $\alpha = 30^0$ (see Fig. 4*a*). Field localization is most prominent in the area comparable with the wavelength for the mode corresponding to diffractive polarization.

### B. Means to control transmission spectrum of CLC with a combined defect

By varying the CLC helix pitch *p* and the angle of light incidence on the sample it is possible to control the transmission spectrum behavior. Increasing the helix pitch, for example by varying the temperature, results in a bandgap shift to the long-wavelength region following the Bragg condition $pn = \lambda$. The short-wavelength peak corresponding to the defect mode disappears; the long-wavelength peak of defect modes corresponding to the right- and left-handed circular polarizations remains. It is essential that there are pitches for which the resonance frequency $\omega_0$ appears to be located close to the short-wavelength boundary of PBG.

An example of the resultant changes in the transmission spectrum of the structure at $\alpha = -30^0$ is illustrated in Fig. 6a. The CLC helix pitch was changed from $p = 275$ nm to $p = 305$ nm, which made the NC resonant frequency $\omega_0$ move close to the short-wavelength boundary of PBG.

Bandgap splitting is observed when the resonance mode is coupled with photonic modes. An extra transparency band induced by diffractive polarization splits off the short-wavelength edge to give rise to a bandgap in the vicinity of $\omega_0$ for waves of both polarizations. The stop band is primarily the result of field absorption in the NC layer. A further increase of the helix pitch forces the resonant frequency $\omega_0$ to shift to the continuous transmission spectrum. The developed resonance situation in this case facilitates formation of an extra bandgap in the transmission spectrum (Fig. 6b).

An alternative approach to realize similar effects is to increase the angle of light incidence on the structure. Then, following the Bragg condition, the bandgap will shift toward the short-wavelength range while the long-wavelength edge of PBG comes closer to the resonant frequency $\omega_0$ of the defect layer.

As mentioned above, high susceptibility to external fields is an essential advantage of CLC over other types of PCs. These fields can be used to control the helix pitch not only of the entire cholesteric but also of part of CLC on the left or on the right of the defect. Consider transmission spectra behavior of the structure when the CLC helix pitch is changed on the right of the defect $p_2$. First we analyze an initial structure of two CLC layers coupled with dielectric defect. Transmission spectrum of the structure is shown in Fig. 3a. There is no twist defect in this case, $\alpha = 0$ and the filling factor is $f = 0$. Figure 7a shows the transmission spectra for two individual CLC layers with various helix pitches, and for a system of these CLC layers separated by a defect.

In the figure we can see that the defect mode occurs in the area where bandgaps of the two CLCs overlap. Our calculations show that the greater the pitch of the right-side helix, the broader the PBG of the system. This fact of PBG broadening and occurrence of a defect mode in the region of overlapping of two scalar PCs was used to obtain a tunable asymmetric filter in a recent experiment [35].

The defect mode is shifted to the long-wavelength range of the spectrum as the right-side helix pitch grows, which results in a lower transmission coefficient in the maximum of the defect mode band. When $p_2$ is not sufficient for bandgaps of the two CLCs to overlap, there is no defect mode realized. Note also the nonmonotonous behavior of the squared modulus maximum of the electric field in a localized defect mode observed as the pitch $p_2$ grows (Fig. 7b). Figure 8 illustrates the behavior of the transmission spectrum and squared field modulus of a long-wavelength defect mode for various pitches $p_2$ when $\alpha = -30°$, $f = 0.02$, other parameters of the system being the same. Broadening of PBG width is observed as the pitch $p_2$ grows (Fig. 8a). Moreover, the maximum of the squared field modulus localized in the defect mode is shifted due to the twist defect (Fig. 8b).

## IV. CONCLUSION

We have studied spectral properties of a cholesteric liquid crystal with a resonant absorbing defect layer of NC in the PC structure combined with a phase jump at the interface between nanocomposite and cholesteric layers. The NC consists of silver nanoballs dispersed in a transparent glass matrix. Numerical analysis of the spectral properties and field distribution in the sample with the described structure has been carried out using the Berreman transfer matrix method. A number of important spectral features of cholesteric with a structural defect have been revealed attributed primarily to the resonance nature of the NC effective permittivity and its essential dependence on the filling factor $f$.

We have studied manifestation of the effect of splitting of defect modes induced for both circular polarizations of incident light in the transmission spectra. Frequency splitting results in a spectral bandgap in the transmission spectrum. The splitting area grows with the volume fraction of nanoballs in the defect layer and can be as large as 50 nm.

We have shown that it is possible to effectively control the transmission spectrum of CLC with a combined defect by varying the angle of incidence of light on the CLC or by applying external fields to vary the helix pitch. There are such angles of incidence or helix pitches that the NC resonant frequency appears to be close to the boundary of the bandgap of a CLC structure, which facilitates emergence of an extra transmission band for waves of both circular polarizations.

Varying the magnitude of phase jump of CLC helix is a nontrivial means of control available only for chiral PCs. Transmission spectrum of the structure under study can be effectively controlled by manipulating the direction and magnitude of the twist defect.

New spectral features arise when the CLC helix pitch is modified on just one side from the defect layer. Increasing the helix pitch leads to widening of PBG of the structure and repositioning of the defect modes. It has been revealed that the electric field maximum of the localized mode varies nonmonotonously with the helix pitch $p_2$.

The undertaken research is important for practical applications as it provides further means to control the transmission spectrum and the degree of field localization in defect modes in CLC with a combined defect as well as improves the application potential of such structures for fabrication of novel optical devices.

## ACKNOWLEDGMENTS

This work was partially supported by grants Nos. 43, 101 and 24.29 of SB RAS; Russian Fund of Fundamental Research, No. 14-02-31248, Project № 3.1276.2014/K under the Government program of the Russian Ministry of Education and Science; Grant of the President of the Russian Federation MK-250.2013.2, NSCT-SB RAS joint project.

**Figure Captions**

Fig. 1. Schematic of CLC with a combination of defects

Fig. 2. Imaginary $\varepsilon''_{mix}$ (dashed curve) and real $\varepsilon'_{mix}$ (solid curve) parts of the effective permittivity of $\varepsilon_{mix}$ versus wavelength. The filling factor $f = 0.02$ (*a*), 0.1 (*b*)

Fig. 3. Transmission spectrum for right- (solid curve) and left-handed (dashed curve) polarizations for α = 0: (a) $f = 0$, (b) $f = 0.02$.

Fig. 4. Transmission spectra of the structure for various α: (a) α > 0, (b) α < 0. Bold curves refer to the right-handed and thin curves to left-handed circular polarizations of light $f = 0.02$.

Fig. 5. Distribution of the squared electric field modulus in defect modes with λ = 388.2 nm (Option 3 in Fig. 4a). (a) Wave with right-handed circular polarization, (b*)* Wave with left-handed circular polarization. $f = 0.02$

Fig.6.Transmission spectrum for various helix pitches. (a) *p* = 305 nm, (b) *p* = 320 nm. Solid and dashed curves refer to the right-handed and left-handed circular polarizations of light, respectively. $f = 0.02$, α = − 30⁰.

Fig. 7. (a) Transmission coefficient for a system of two cholesterics with a defect (solid curve) and for each of the cholesterics (dashed curve) with the pitch on the left of the defect $p_1 = 275$ nm (dotted curve) and $p_2 = 290$ nm (dotted curve). (b) Maximum squared modulus of the electric field of the defect mode as a function of the pitch $p_2$. The remaining parameters are the same as in Fig.3a.

Fig. 8 (a) Transmission spectra for various $p_2$: dash-and-dotted curve $p_2 = 275$ nm, dashed curve $p_2 = 290$ nm, solid curve $p_2 = 305$ nm, α = − 30⁰. (b) Maximum of the squared electric field modulus of the long-wavelength defect mode as a function of the pitch $p_2$, α = − 30⁰. Bold curves refer to the right-handed and thin curves to the left-handed circular polarizations of light, $f = 0.02$

**Figures**

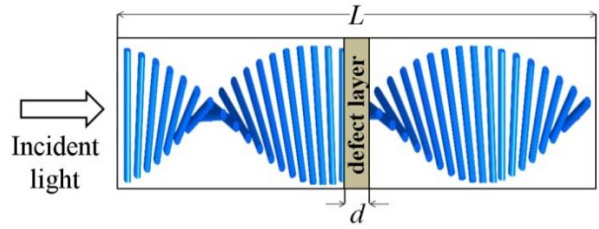

Fig. 1. Schematic of CLC with a combination of defects

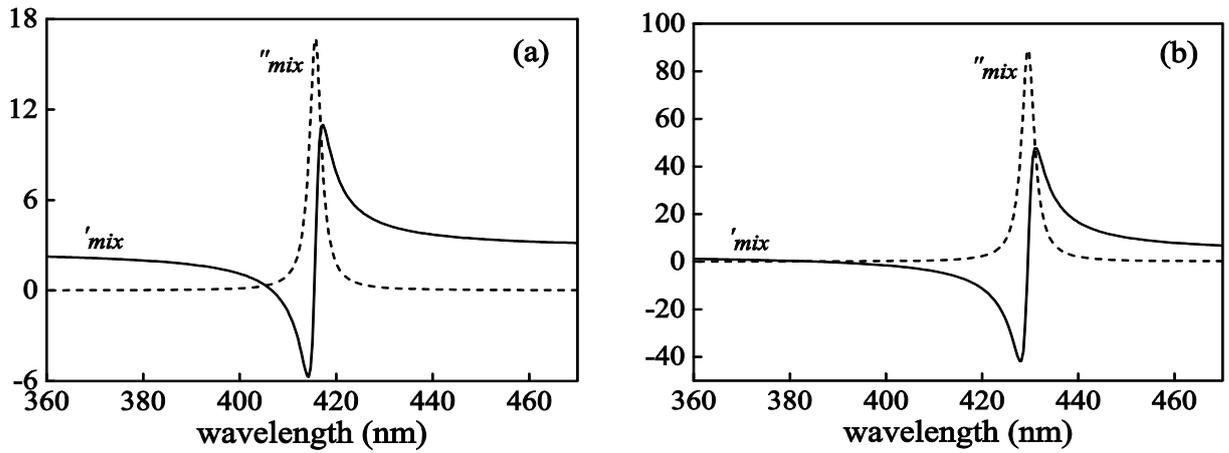

Fig. 2. Imaginary $\varepsilon''_{mix}$ (dashed curve) and real $\varepsilon'_{mix}$ (solid curve) parts of the effective permittivity of $\varepsilon_{mix}$ versus wavelength. The filling factor $f = 0.02$ (a), 0.1 (b)

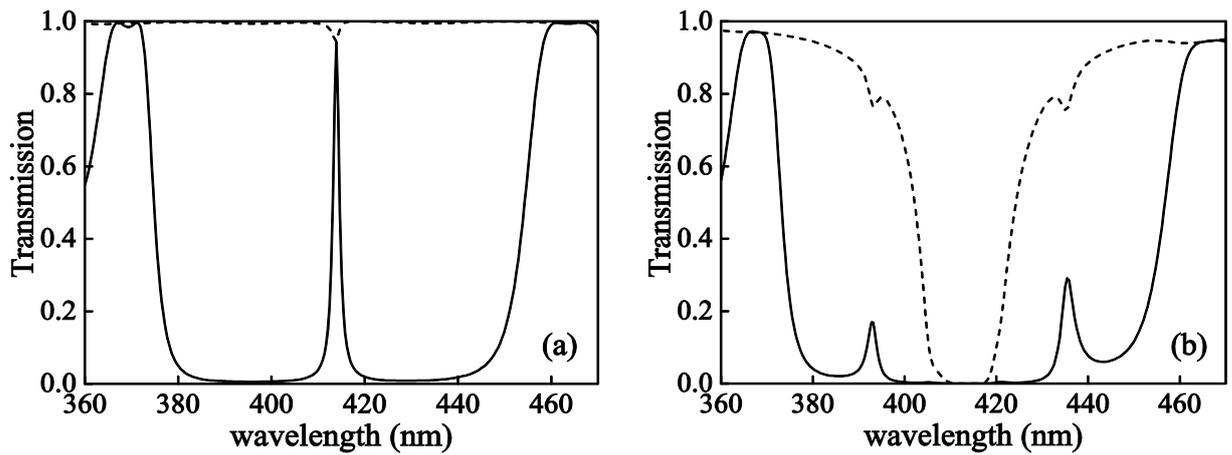

Fig. 3. Transmission spectrum for right- (solid curve) and left-handed (dashed curve) polarizations for $\alpha = 0$: (a) $f = 0$, (b) $f = 0.02$.

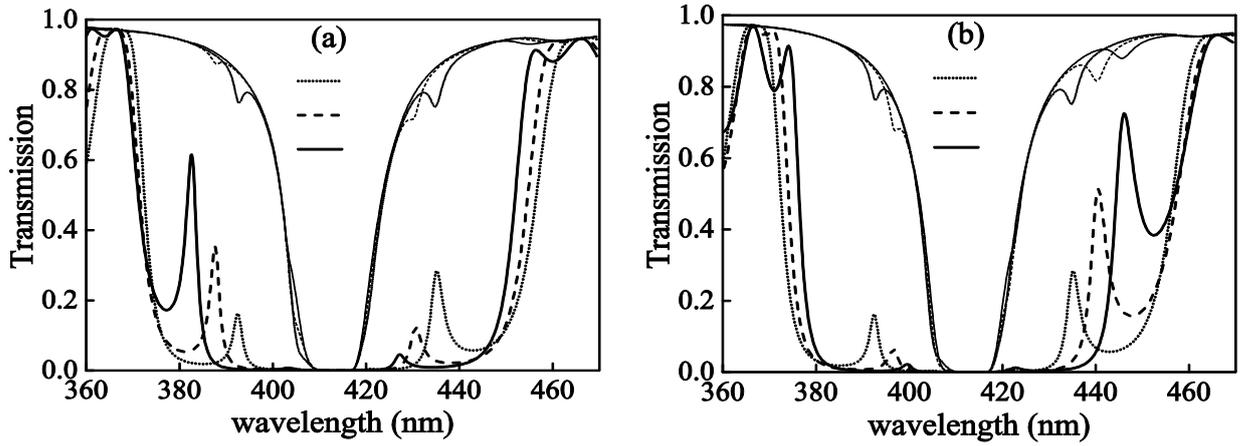

Fig. 4. Transmission spectra of the structure for various α: (a) α > 0, (b) α < 0. Bold curves refer to the right-handed and thin curves to left-handed circular polarizations of light *f* = 0.02.

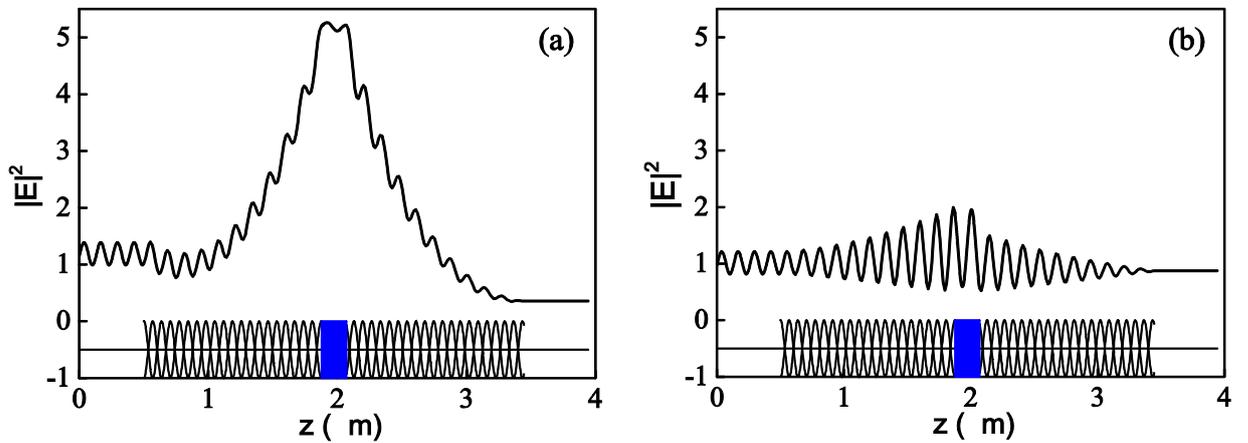

Fig. 5. Distribution of the squared electric field modulus in defect modes with λ = 388 nm (Option 3 in Fig. 4a). (a) Wave with right-handed circular polarization, (b) Wave with left-handed circular polarization. *f* = 0.02

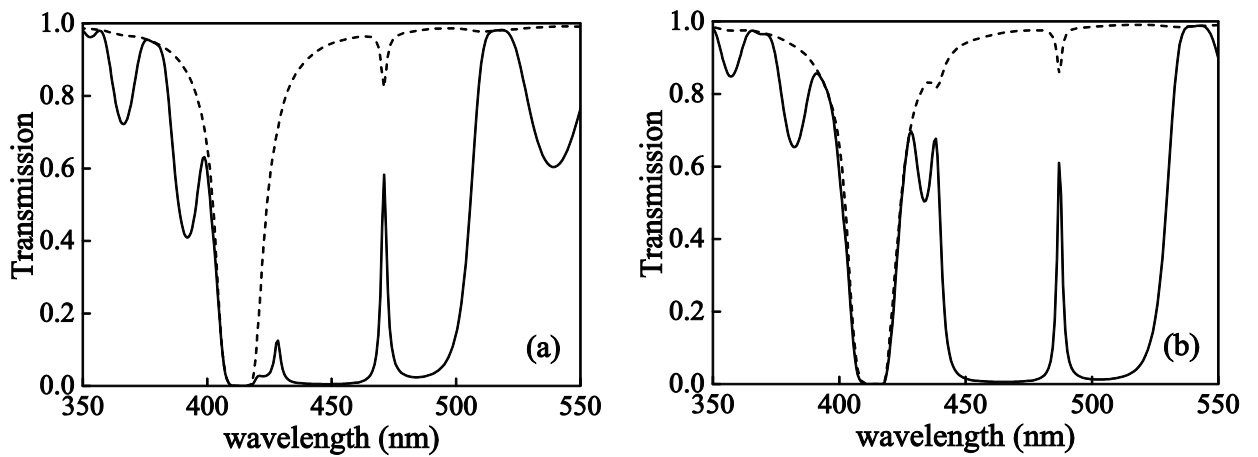

Fig. 6. Transmission spectrum for various helix pitches. (a) *p* = 305 nm, (b) *p* = 320 nm. Solid and dashed curves refer to the right-handed and left-handed circular polarizations of light, respectively. *f* = 0.02, α = − 30⁰.

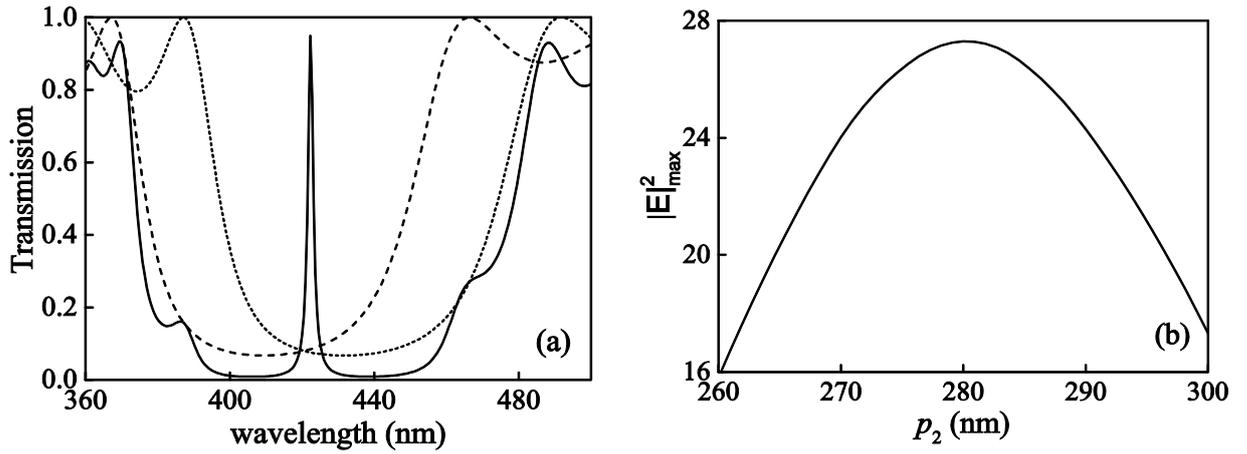

Fig. 7. (a) Transmission coefficient for a system of two cholesterics with a defect (solid curve) and for each of the cholesterics (dashed curve) with the pitch on the left of the defect $p_1 = 275$ nm (dotted curve) and $p_2 = 290$ nm (dotted curve). (b) Maximum squared modulus of the electric field of the defect mode as a function of the pitch $p_2$. The remaining parameters are the same as in Fig.3a.

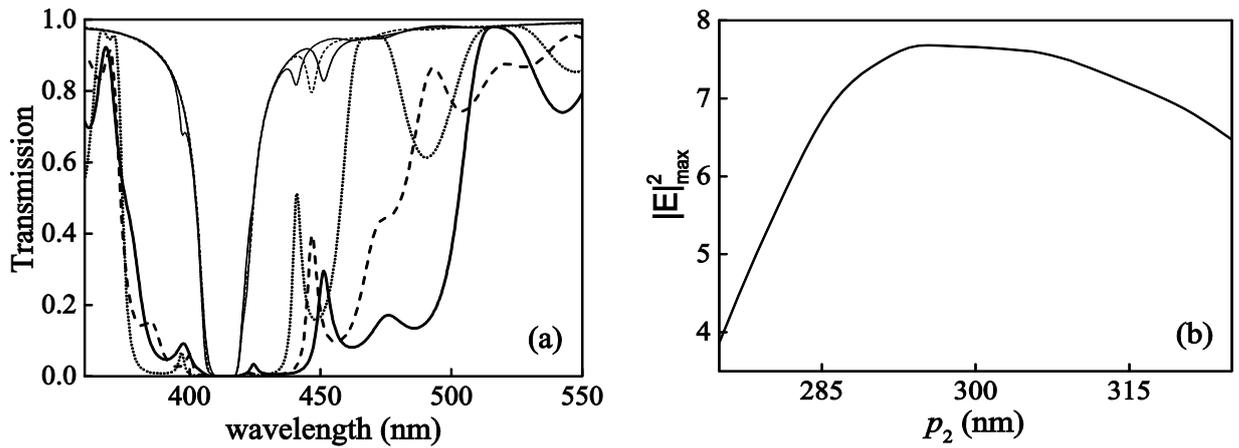

Fig. 8 (a) Transmission spectra for various $p_2$: dash-and-dotted curve $p_2 = 275$ nm, dashed curve $p_2 = 290$ nm, solid curve $p_2 = 305$ nm, $\alpha = -30^0$. (b) Maximum of the squared electric field modulus of the long-wavelength defect mode as a function of the pitch $p_2$, $\alpha = -30^0$. Bold curves refer to the right-handed and thin curves to the left-handed circular polarizations of light, $f = 0.02$